\RequirePackage{fix-cm}
\documentclass[reprint,aps,groupedaddress,nofootinbib,showkeys]{revtex4}

\RequirePackage{graphicx}
\usepackage{mathbbol}
\usepackage{graphicx}
\usepackage{amsmath}
\usepackage{dcolumn}
\usepackage[utf8]{inputenc}
\usepackage{bm}
\usepackage{xcolor}
\usepackage{color}
\usepackage[caption=false]{subfig}
\usepackage{multirow}
\usepackage{newtxtext,newtxmath}

\begin{document}

\title{A note on the exact solution for the magnetic field in a solenoid}

\author{G. Dattoli$^{1},$ E. Di Palma$^{1},$    E. Sabia$^{1}$}

%

\affiliation{$^{1}$ENEA - Centro Ricerche Frascati, Via Enrico Fermi 45, 00044, Frascati, Rome, Italy}

\begin{abstract}
We comment on a recent paper regarding the derivation of the magnetic field components of a solenoid in analytical form by proposing a different and simpler method
\end{abstract}

\keywords{Solenoid, magnetic field, Maxwell equations, elliptic integrals, Bessel functions, hypergeometric functions.}

\maketitle

\section{Introduction}

In a recent interesting paper \cite{Behtouei}, an analytical solution for the evaluation of the field in a solenoid has been proposed. In this note we make a step further by proposing an alternative solution, which has the only merit of simplifying the proposed solution procedure.

In the quoted paper it has been shown that the field distribution can be specified through the following two elliptic-like integrals \cite{Andrew}

\begin{equation}
\begin{array}{l}
{i_{1} (\xi )=\displaystyle \int _{0}^{\pi }\dfrac{1}{\left[1-\xi \, \cos (\phi )\right]^{\frac{3}{2} } }  d\phi ,} \\ \\
{i_{2} (\xi )=\displaystyle \int _{0}^{\pi }\dfrac{\cos (\phi )}{\left[1-\xi \, \cos (\phi )\right]^{\frac{3}{2} } }  d\phi }
\end{array}
\label{GrindEQ__1_}
\end{equation}

In order to get an analytical expression, we cast $i_{1} (\xi )$ in the form

\begin{equation}
\label{GrindEQ__2_}
i_{1} (\xi )=\frac{1}{\Gamma \left(\frac{3}{2} \right)} \int _{0}^{\pi }\left[\int _{0}^{\infty }e^{-s} s^{\frac{1}{2} }  ds\right] e^{s\xi \cos (\phi )} d\phi  \end{equation}

Obtained after exploiting the Laplace transform identity \cite{Andrew,Babusci}

\begin{equation}
\label{GrindEQ__3_}
a^{-\nu } =\frac{1}{\Gamma (\nu )} \int _{0}^{\pi }e^{-sa} s^{\nu -1}  ds
\end{equation}

The use of the 0-th order modified Bessel function integral representation \cite{Abramowitz}

\begin{equation}
\label{GrindEQ__4_} I_{n} (x)=\frac{1}{\pi } \int _{0}^{\infty }e^{x\cos (\phi )} \cos (n\, \phi )d\phi
\end{equation}

eventually yields

\begin{equation}
\label{GrindEQ__5a_}
i_{1} (\xi )=\frac{\pi }{\Gamma \left(\frac{3}{2} \right)} \int _{0}^{\infty }\left[e^{-s} I_{0} (\xi \, s)s^{\frac{1}{2} } \right] ds.
\end{equation}

The use of the series expansion for the modified Bessel \cite{Andrew}

\begin{equation}
I_{\nu } (z)=\sum _{r=0}^{\infty }\frac{\left(\frac{z}{2} \right)^{2r+\nu } }{r!\Gamma (r+\nu +1)} \;\; \mbox{with}\;\;\nu \in R,
\label{GrindEQ__5b_}
\end{equation}

and of the integral  form of the Gamma function \cite{Andrew}

\begin{equation}
\begin{array}{l}
{\Gamma (\nu )=\int _{0}^{\infty }e^{-t}  t^{\nu -1} dt \;\; \mbox{with}\;\;  \nu \in R}, \\ \\
{\Gamma (n+1)=n! \;\;  \mbox{with}\;\;  n \in N} ,
\end{array}
\label{GrindEQ__5c_}
\end{equation}

Yields the following straightforward integration of eq. \ref{GrindEQ__5a_}

\begin{equation}
\label{GrindEQ__6_} i_{1} (\xi )=2\sqrt{\pi } f_{0} (\xi )
\end{equation}

Where $f_{0} (\xi )$ is a Bessel-like function defined by the series

\begin{equation}
\label{GrindEQ__7_}
\begin{array}{l}
{f_{0} (\xi )=\displaystyle \sum _{r=0}^{\infty }\dfrac{\Gamma \left(2\, r+\dfrac{3}{2} \right)}{r!^{2} }  \left(\dfrac{\xi }{2} \right)^{2\, r} ,} \\ \\
{\left|\xi \right|<1} ,
\end{array}
\end{equation}

In Figs. \ref{GrindEQ__1_} we have reported a comparison between analytical and numerical integration along with the relative error

\begin{figure}[h]
\centering
\includegraphics[width=.8\textwidth]{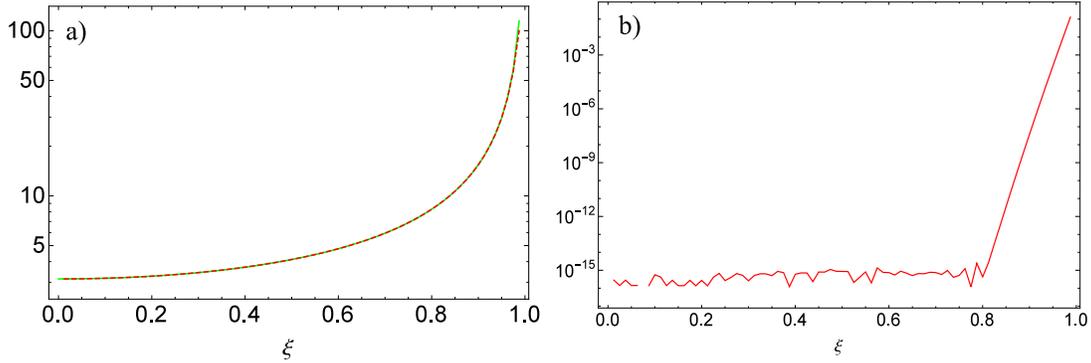}
\caption{a) Comparison between numerical (continuous line) and analytical (dashed-line) solutions; b) relative error. }
\label{GrindEQ__1_}
\end{figure}

The extension of the same method to the analytical evaluation of the second integral yields
\begin{equation}
\label{GrindEQ__8_}
\begin{array}{l}
{i_{1} (\xi )=2\sqrt{\pi } f_{1} (\xi )}, \\ \\
{f_{1} (\xi )=\displaystyle \sum _{r=0}^{\infty }\dfrac{\Gamma \left(2\, r+\dfrac{5}{2} \right)}{r!(r+1)!}  \left(\dfrac{\xi }{2} \right)^{2\, r+1} ,} \\ \\
{\left|\xi \right|<1}.
\end{array}
\end{equation}

The comparison between analytical and numerical integration is given in Fig. \ref{GrindEQ__2_}
\begin{figure}[h]
\centering
\includegraphics[width=.8\textwidth]{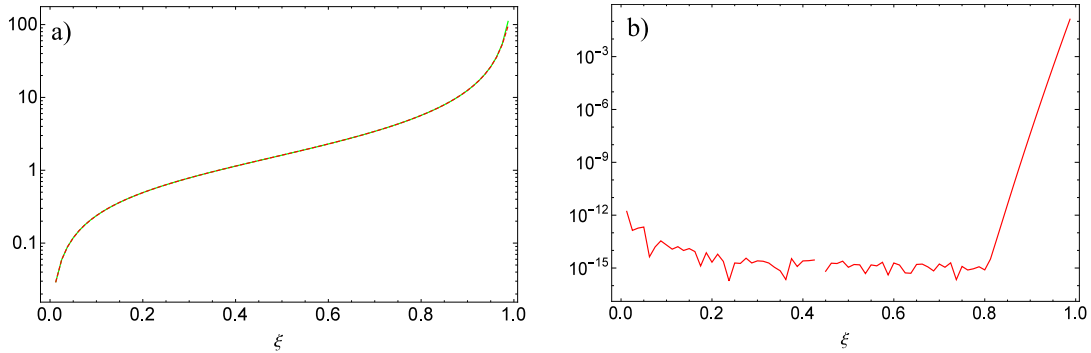}
\caption{Comparison between numerical (continuous-line) and analytical (dashed-line) solutions same as Fig. \eqref{GrindEQ__1_}. }
\label{GrindEQ__2_}
\end{figure}

The method we have envisaged is quite straightforward and requires fairly elementary means.

In ref. \eqref{GrindEQ__1_} a more elaborated technique, requiring fractional differ-integrals and complex integration has been employed to get the explicit form of the integrals in terms of hyper-geometric series, namely

\begin{equation}
\label{GrindEQ__9_}
\begin{array}{l}
{i_{1} (\xi )=\dfrac{\pi }{(1+\xi )^{3/2 } } {}_{2} F_{1} \left(\frac{1}{2} ,\, \frac{3}{2} ;\, 1,\frac{2\xi }{1+\xi } \right),} \\ \\
{i_{2} (\xi )=\dfrac{\pi }{(1+\xi )^{3/2 } } \left( {}_{2} F_{1} \left(\frac{3}{2} ,\, \frac{3}{2} ;\, 2,\frac{2\xi }{1+\xi }\right)-{}_{2} F_{1} \left(\frac{1}{2} ,\, \frac{3}{2} ;\, 1,\frac{2\xi }{1+\xi } \right)\right)}
\end{array}
\end{equation}

where

\begin{equation}
\label{GrindEQ__10_}
\begin{array}{l}
{F_{1} \left(a,\, b;\, c,z\right)=\displaystyle \sum _{n=0}^{\infty }\dfrac{(a)_{n} \left(b\right)_{n} }{\left(c\right)_{n} }  \dfrac{z^{n} }{n!} ,} \\ \\
{(k)_{n} =\left\{ \begin{array}
{c} {1} \\ \\
{a(a+1)...(a+n-1)} \end{array}\; \;
\begin{array}{c}
{n=0} \\ \\
{n>0}
\end{array} \right. }
\end{array}
\end{equation}

In fig. \ref{GrindEQ__3_} we provide a comparison between the ``Bessel'' and hyper-geometric series.

The differences are evident for larger values of $\xi \, (\left|\xi \right|<1)$, where the convergence of the hyper-geometric series is slower. A better agreement could be obtained using different representation of the hyper-geometric functions.

\begin{figure}[h]
\centering
\includegraphics[width=.8\textwidth]{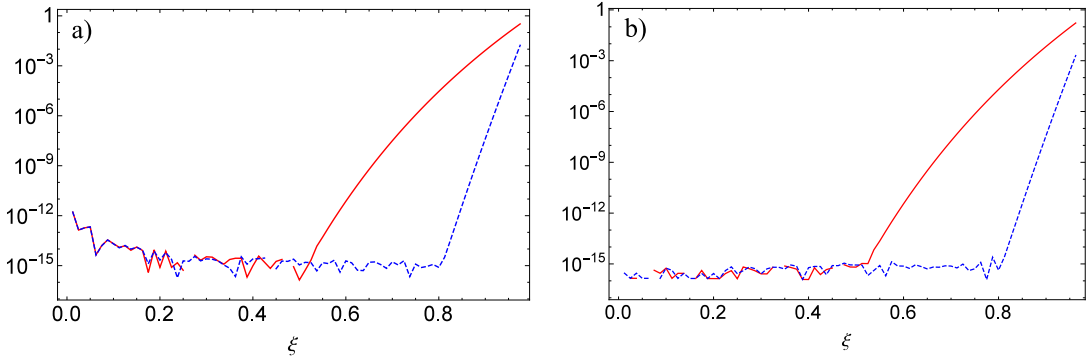}
\caption{a) Relative error between numerical integration of $i_{1} (\xi )$ and Bessel (dot line), hyper-geometric (continuous line) solutions; b) Same as a) for $i_{2} (\xi )$. }
\label{GrindEQ__3_}
\end{figure}

In order to appreciate the relevance of the variable $\xi$ to the computation of the magnetic field of an actual solenoid we provide, in Fig. \ref{GrindEQ__4_}, the relevant geometry. We note that, in terms of the solenoid parameters $\xi$ reads
\begin{equation}
\label{GrindEQ__11_}
\begin{array}{l}
{\xi (\eta ,\, \zeta )=\dfrac{2\eta }{1+\eta ^{2} +\zeta ^{2} } ,\;\; \eta =\dfrac{r}{R} ,\;\; \zeta =\dfrac{z}{R} }, \\ \\
{r\equiv {\rm distance\; from\; the\; solenoid\; axis}} \\ \\
 {R\equiv {\rm solenoid\; radius}}.
\end{array}
\end{equation}

\begin{figure}[h]
\centering
\includegraphics[width=.6\textwidth]{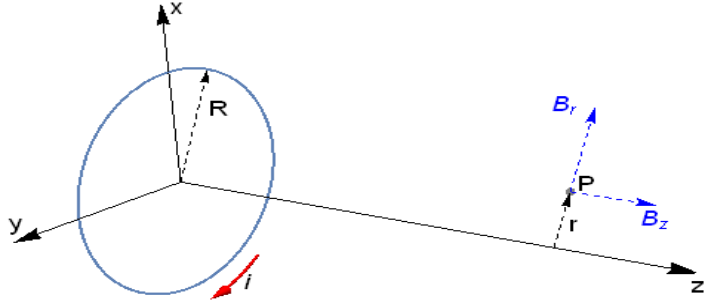}
\caption{Solenoid geometry and field components.}
\label{GrindEQ__4_}
\end{figure}

\begin{figure}[h]
\centering
\includegraphics[width=.4\textwidth]{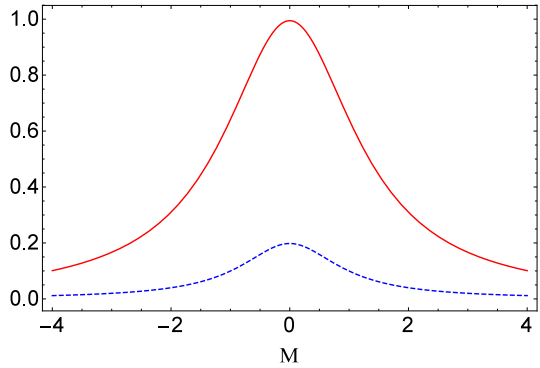}
\caption{$\xi$ variable vs. $\zeta$ for different values of $\eta$: $0.9$ continuous-line and $0.1$ dot-line.}
\label{GrindEQ__5_}
\end{figure}

In Fig. \eqref{GrindEQ__5_} we have reported $\xi$ vs $\zeta$ for different values of $\eta$.

We expect deviation between numerical and analytical solutions for increasing values of $\eta$, which measures the distance from the solenoid axis the distance. Recalling that the profile of the radial field vs$\xi$ is provided by
\begin{equation}
\label{GrindEQ__12_}
B_{r} (\eta ,\zeta )\propto \left(\dfrac{\xi (\eta ,\zeta )}{\eta } \right)^{3/2} i_{2} (\xi (\eta ,\zeta )),
\end{equation}

and the comparison between numerical and Bessel like/hyper-geometric analytic is provided in Fig. \ref{GrindEQ__6_}. The disagreement is due to the poor convergence of the series for large values of $\eta$, it should be noted that complete agreement is restored for $\eta \le 0.7$

\begin{figure}[h]
\centering
\includegraphics[width=.5\textwidth]{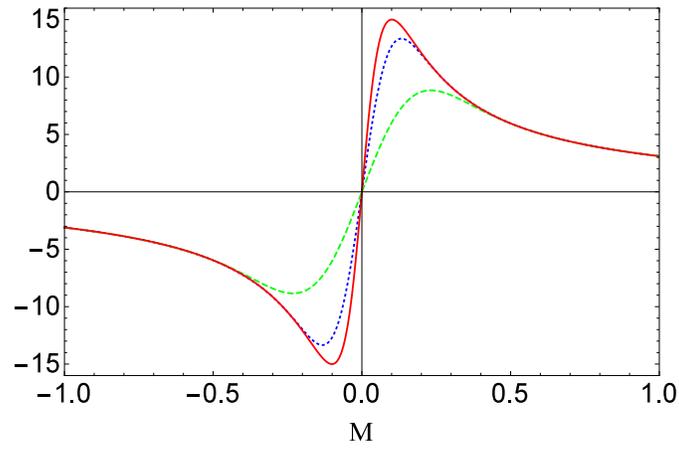}
\caption{Profile of the radial magnetic field vs. $\zeta $ for $\eta =0.9$. a) Hyper-Geometric Analytical Solution (dashed-line), b) Bessel-like analytic solution (dotted-line), c) numerical integration (continuous-line)}
\label{GrindEQ__6_}
\end{figure}

Before concluding, we make a comparison with a more realistic computation employing CST$^\circledR$ microwave studio \cite{CST}. The CST simulation has been performed  for $\eta =4\cdot 10^{-3}$ using either a coil, with a section of $3\;mm$, and an ideal wire. In Fig. \eqref{GrindEQ__7_} we provide the comparison with the Bessel like function and find that the wire solution compares better with the analytical counterpart.

In fig. \eqref{GrindEQ__8_} we have reported the same comparison (for the case of the coil only) and we find a significant disagreement around the maxima, due to the poor convergence properties of the Bessel series for large $\eta$ (and hence $\xi$) values.

\begin{figure}[h]
\centering
\includegraphics[width=.8\textwidth]{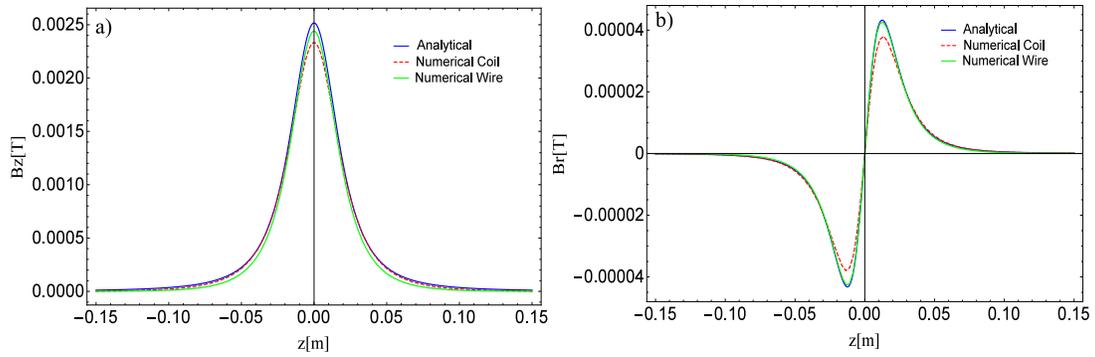}
\caption{Comparison among analytical and numerical solution (present method) for longitudinal and radial (b) magnetic fields generated by a simple wire and a coil with a cross section $t=3\; mm$ and depth  $0.0001$ for $\eta=0.004$.}
\label{GrindEQ__7_}
\end{figure}

\begin{figure}[h]
\centering
\includegraphics[width=.8\textwidth]{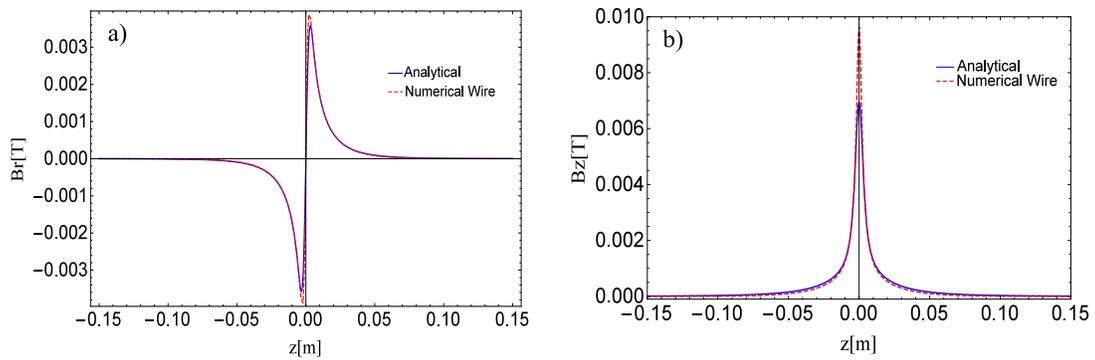}
\caption{Same as Fig. \eqref{GrindEQ__7_} for the field generated  by a wire $(\eta =0.9)$.}
\label{GrindEQ__8_}
\end{figure}

This note has addressed a minor problem regarding the possibility of simplifying the mathematical formalism underlying the derivation of the magnetic fields of a finite length solenoid. Further consequences will be drawn in a forthcoming investigation.

\bibliographystyle{jpp}

\end{document}